\documentclass[a4paper,fleqn,usenatbib,useAMS]{mnras}

\usepackage[T1]{fontenc}
\usepackage{ae,aecompl}
\usepackage{natbib}

\usepackage{graphicx}	
\usepackage{amsmath}	
\usepackage{amssymb}	
\usepackage{multicol}        
\usepackage{bm}		
\usepackage{pdflscape}	
\usepackage{booktabs}
\usepackage{caption}
\usepackage{mathtools}
\usepackage{makecell}

\newcommand{\fesc}{\ensuremath{f_{\rm esc}}}
\newcommand{\fout}{\ensuremath{\langle f_{900}/f_{1500}\rangle_{\rm out}}}
\newcommand{\fobs}{\ensuremath{\langle f_{900}/f_{1500}\rangle_{\rm obs}}}
\newcommand{\ziii}{\ensuremath{z\sim3}}
\newcommand{\wlya}{$W_{\lambda}$(Ly$\rm \alpha$)}

\newcommand{\luv}{\ensuremath{L_{\rm UV}}}

\newcommand{\fesca}{\ensuremath{f_{\rm esc,abs}}}
\newcommand{\zspec}{\ensuremath{z_{\rm sys}}}
\newcommand{\sfrd}{\ensuremath{\Sigma_{\rm SFR}}}

\newcommand{\vjh}{\ensuremath{V_{606}J_{125}H_{160}}}
\newcommand{\hstj}{\ensuremath{J_{125}}}
\newcommand{\hstv}{\ensuremath{V_{606}}}
\newcommand{\hsth}{\ensuremath{H_{160}}}
\newcommand{\mstar}{\ensuremath{\rm M_{\rm *}}}
\newcommand{\msun}{\ensuremath{M_{\odot}}}

\newcommand{\lya}{Ly$\rm \alpha$}
\newcommand{\oiii}{[O~\textsc{iii}]\ensuremath{\lambda\lambda4959,5007}}
\newcommand{\oii}{[O~\textsc{ii}]\ensuremath{\lambda\lambda3726,3729}}
\newcommand{\hb}{H$\rm \beta$}

\title[LyC escape and galaxy properties]{The connection between the escape of ionizing radiation and galaxy properties at $z\sim3$ in the Keck Lyman Continuum Spectroscopic Survey}

\author[Pahl et al.]{Anthony J. Pahl,$^{1}$\thanks{Contact e-mail: \href{mailto:pahl@astro.ucla.edu}{pahl@astro.ucla.edu}}
Alice Shapley,$^{1}$
Charles C. Steidel,$^{2}$
Naveen A. Reddy,$^{3}$\newauthor
Yuguang Chen,$^{2,4}$ 
Gwen C. Rudie,$^5$
and Allison L. Strom$^6$
	\\
	$^{1}$Department of Physics and Astronomy, University of California, Los Angeles, CA 90095, USA\\
	$^{2}$Cahill Center for Astronomy and Astrophysics, California Institute of Technology, MC249-17, Pasadena, CA 91125, USA\\
	$^{3}$Department of Physics and Astronomy, University of California Riverside, Riverside, CA 92521, USA\\
	$^{4}$Department of Physics and Astronomy, University of California Davis, 1 Shields Avenue, Davis, CA 95616, USA\\
	$^{5}$The Observatories of the Carnegie Institution for Science, 813 Santa Barbara Street, Pasadena, CA 91101, USA\\
	$^{6}$Center for Interdisciplinary Exploration \& Research in Astrophysics (CIERA) and Department of Physics \& Astronomy,\\ Northwestern University, Evanston, IL, 60208, USA 
}

\date{}

\pubyear{}

\begin{document}
		
	\label{firstpage}
	\pagerange{\pageref{firstpage}--\pageref{lastpage}}
	\maketitle
	
\begin{abstract}
	The connection between the escape fraction of ionizing radiation (\fesc{}) and the properties of galaxies, such as stellar mass (\mstar{}), age, star-formation rate (SFR), and dust content, are key inputs for reionization models, but many of these relationships remain untested at high redshift. We present an analysis of a sample of 96 \ziii{} galaxies from the Keck Lyman Continuum Spectroscopic Survey (KLCS). These galaxies have both sensitive Keck/LRIS spectroscopic measurements of the Lyman continuum (LyC) region, and multi-band photometry that places constraints on stellar population parameters. We construct composite spectra from subsamples binned as a function of galaxy property and quantify the ionizing-photon escape for each composite. We find a significant anti-correlation between \fesc{} and \mstar{}, consistent with predictions from cosmological zoom-in simulations. We also find significant anti-correlation between \fesc{} and E(B-V), encoding the underlying physics of LyC escape in our sample. We also find no significant correlation between \fesc{} and either stellar age or specific SFR (=SFR/\mstar{}), challenging interpretations that synchronize recent star formation and favorable conditions for ionizing escape. The galaxy properties now shown to correlate with \fesc{} in the KLCS are Ly$\alpha$ equivalent width, UV Luminosity, \mstar{}, SFR, and E(B-V), but not age or sSFR. To date, this is the most comprehensive analysis of galaxy properties and LyC escape at high redshift, and will be used to guide future models and observations of the reionization epoch.
\end{abstract}


\defcitealias{naiduRapidReionizationOligarchs2020}{N20}

\section{Introduction} \label{sec:intro}

Reionization is a significant phase transition in the Universe, during which hydrogen in the intergalactic medium (IGM) transitions from neutral to ionized. This transformation appears to end at $z\sim6$ \citep{fanObservationalConstraintsCosmic2006}, but we still lack a comprehensive description of the physical processes responsible for the reionization history. At $z>6$, the ionizing background of the Universe is thought to be dominated by ionizing photons produced in O/B stars in star-forming galaxies \citep{bouwensREIONIZATIONPLANCKDERIVED2015,parsaNoEvidenceSignificant2018}.
Different models of reionization attempt to use integrated measurements of galaxy populations and independent constraints on the IGM hydrogen neutral fraction to infer the comprehensive evolution of reionization from the early Universe to $z\sim6$, but can arrive at remarkably different answers. Reionization may end ``late," such that neutral fractions remain at ~$\gtrsim90\%$ until $z\sim8$ \citep[e.g.,][]{naiduRapidReionizationOligarchs2020}, or reionization may be ``gradual," such that the neutral fraction decreases slowly from $z\sim12$ until $z\sim6$ \citep[e.g.,][]{finkelsteinConditionsReionizingUniverse2019}.

To draw conclusions about the evolution of the neutral fraction from observations of galaxies, one must attempt to understand the ionizing emissivity of galaxies as a function of cosmic time. This quantity is commonly parameterized as a function of three variables: the UV luminosity function ($\rho_{\rm UV}$), the ionizing photon production efficiency ($\xi_{\rm ion}$), and the fraction of ionizing luminosity that escapes the interstellar and circumgalactic medium (ISM and CGM) and proceed to ionize the IGM (\fesc{}) \citep{robertsonCosmicReionizationEarly2015}. While constraints are available for both $\rho_{\rm UV}$ and $\xi_{\rm ion}$ well into the epoch of reionization \citep{madauCosmicStarformationHistory2014,starkSpectroscopicDetectionIV2015,starkLyaIIIEmission2017}, \fesc{} is uniquely difficult to ascertain in the early Universe. Estimating \fesc{} requires direct observations of the ionizing radiation from galaxies in the Lyman continuum (LyC) spectral region. The transmission through the general IGM of LyC photons escaping a galaxy depends sensitively on emission redshift and decreases rapidly beyond $z\sim3.5$ \citep{vanzellaDetectionIonizingRadiation2012}. This drop off is due to LyC absorption from trace amounts of H~\textsc{i} and makes direct determinations of \fesc{} impossible during the epoch of reionization itself.

Models of reionization are distinguished by their assumptions about \fesc{}. \citet{finkelsteinConditionsReionizingUniverse2019} present a ``democratic" model for reionization that assumes the process is driven by faint sources with high \fesc{} values. In contrast, the ``oligarchical" model of \citet{naiduRapidReionizationOligarchs2020} concludes that massive, luminous ($M_{\rm UV}<-18$ and log(\mstar{}/\msun{})$>8$) galaxies provide the bulk of ionizing photons during reionization. For testing assumptions of \fesc{} during reionization, $z\sim3-4$ galaxies provide an essential laboratory, and can discern the fundamental properties that govern \fesc{} at the highest redshifts these measurements can be made. Critically, these galaxies may be closer analogs to reionization era galaxies than those in the local universe \citep[e.g.,][]{fluryLowredshiftLymanContinuum2022,fluryLowredshiftLymanContinuum2022a}.

A number of LyC observational surveys at $z\sim3-4$ have attempted to measure average \fesc{} values and potential correlations between \fesc{} and the properties of galaxies. Success has been found by stacking deep observations of the LyC either photometrically \citep[e.g., ][]{begleyVANDELSSurveyMeasurement2022} or spectroscopically \citep{marchiNewConstraintsAverage2017}. Here we focus on the Keck Lyman Spectroscopic (KLCS) survey, which included deep Keck/Low Resolution Imaging Spectrometer  \citep[LRIS;][]{okeKeckLowResolutionImaging1995,steidelSurveyStarForming2004} spectra of Lyman break galaxies (LBGs) at $z\sim3$ \citep{steidelKeckLymanContinuum2018}. \citet{steidelKeckLymanContinuum2018} reported an average $\fesc=0.09\pm0.01$ from 124 LBGs, estimated by stacking their rest-UV spectra with coverage of the LyC region. After careful treatment of line-of-sight contamination using \textit{HST} imaging, the average KLCS \fesc{} was corrected to $0.06\pm0.01$ upon removal of four apparently-contaminated galaxies from the sample \citep{pahlUncontaminatedMeasurementEscaping2021}.
Galaxy properties correlated with inferred \fesc{} in the KLCS are determined to be Ly$\alpha$ equivalent width (\wlya{}) and UV luminosity (\luv{}), such that galaxies with stronger Ly$\alpha$ emission and lower \luv{} luminosities tend to have higher \fesc{} \citep{steidelKeckLymanContinuum2018,pahlUncontaminatedMeasurementEscaping2021}. In the 2018 paper, we explained why \wlya{} is more fundamental in its correlation with \fesc{}, as \wlya{} is modulated by the neutral gas covering fraction of a galaxy \citep{reddyConnectionReddeningGas2016,gazagnesOriginEscapeLyman2020}, which similarly modulates the escape of ionizing radiation. Trends between \fesc{}, \wlya{}, and \luv{} have also been recovered in complementary LyC surveys at \ziii{} \citep{marchiNewConstraintsAverage2017,marchiLyaLymanContinuumConnection2018,begleyVANDELSSurveyMeasurement2022}, but \luv{} and \wlya{} ultimately represent a limited parameter space from which to construct a comprehensive picture of LyC escape in star-forming galaxies. 

Promising indirect indicators of \fesc{} may surface from the feedback of star formation and its effect on the ISM and CGM of a galaxy. Cosmological zoom-in simulations coupled with radiative transfer calculations indicate that feedback from recent, dense star-formation can induce favorable channels in the ISM and CGM that allow ionizing photons to escape \citep{maNoMissingPhotons2020}. Understanding \fesc{} as a function of the surface density of star-formation (\sfrd{}), stellar age, or specific star-formation rate would allow observational comparison to these simulations, and empirically connect the history of star formation to \fesc{}. Additionally, dust attenuation is intricately linked to the neutral gas covering fraction in the ISM and CGM, but the relationship between \fesc{} and E(B-V) at \ziii{} has thus far only been investigated using rest-UV observations \citep{reddySPECTROSCOPICMEASUREMENTSFARULTRAVIOLET2016,reddyConnectionReddeningGas2016,steidelKeckLymanContinuum2018}. Thanks to multi-band photometry available for the KLCS, which can constrain stellar population parameters, we can explore these relationships at high redshift, many for the first time.

In \citet{pahlSearchingConnectionIonizingphoton2022}, we began by examining 35 galaxies from the KLCS that were covered by \textit{HST} imaging, enabling measurements of rest-UV sizes. Together with SFR estimates from fits to multi-band photometry, we measured \sfrd{} and attempted to constrain \fesc{} vs. \sfrd{}. We ultimately determined that the limited KLCS subsample with \textit{HST} imaging was too small and unrepresentative to determine trends with \fesc{} and galaxy property. 
In the present work, we extend the analysis of \citet{pahlSearchingConnectionIonizingphoton2022} by examining SED-modeled measurements of stellar mass (\mstar{}), E(B-V), stellar age, and SFR instead, which allow nearly the entire KLCS sample to be utilized. By performing stacking of rest-UV spectra as a function of galaxy property, we investigate the dependence of \fesc{} on these galaxy properties. Significant correlations will test existing reionization models and strongly inform future ones.

We organize the paper as follows: in Section \ref{sec:methods}, we review the spectroscopic observations of the KLCS sample and its ancillary photometric measurements, and provide an overview of the SED and spectral fitting methodology. In Section \ref{sec:res}, we present the SED-modeled parameters for individual galaxies and estimates of ionizing escape from binned subsamples. In Section \ref{sec:disc}, we explore similar observational analyses from the literature, connections to cosmological zoom-in simulations, and implications for reionization. We summarize our main conclusions in Section \ref{sec:summary}.

Throughout this paper, we adopt a standard $\Lambda$CDM cosmology with $\Omega_m$ = 0.3, $\Omega_{\Lambda}$ = 0.7 and $H_0$ = 70 $\textrm{km\,s}^{-1}\textrm{Mpc}^{-1}$. The {\fesc} values reported in this paper are absolute escape fractions, equivalent to {\fesca} in \citet{steidelKeckLymanContinuum2018}, and defined as the fraction of all H-ionizing photons produced within a  galaxy that escapes into the IGM. We also employ the AB magnitude system \citep{okeSecondaryStandardStars1983}.

\section{Sample and Methodology} \label{sec:methods}

In order to understand how ionizing photon escape is tied to measurable characteristics of galaxy stellar populations or the spatial distribution of the interstellar gas, we require integrated photometric measurements that sample a wide wavelength baseline as well as direct constraints on the LyC emission. Both types of measurements are available for KLCS galaxies. In this section, we outline the data included in our analysis, featuring an overview of the KLCS sample, associated spectra, and the multi-band photometry available for KLCS galaxies. We explain the methodology of SED fitting to determine galaxy properties and spectral modeling to estimate parameterizations of ionizing-photon escape such as \fesc{}.


\subsection{Uncontaminated KLCS}

The primary goal of the KLCS was to examine the hydrogen-ionizing spectra of star-forming galaxies at \ziii{} \citep{steidelKeckLymanContinuum2018}. 
To this end, 136 galaxies were observed with LRIS on the Keck I telescope on Mauna Kea, Hawai'i. Each object was observed for 8.2 hours at minimum. These observations began in 2006 and were concluded in 2008. Of 139 targets, 13 galaxies were removed due to either instrumental defects or spectroscopic evidence of contamination by foreground galaxies. The final sample presented in \citet{steidelKeckLymanContinuum2018} numbered 124 galaxies. Of these, 15 galaxies apparently had significant flux density in the LyC spectral region, defined as having $f_{900}>3\sigma_{900}$, where $f_{900}$ is the average flux density between $880-910$\AA{} in the rest frame, and $\sigma_{900}$ is the standard deviation of flux densities in the same spectral region. Objects meeting this criteria were defined as individual LyC ``detections," with the remaining 109 galaxies labeled as LyC ``non-detections."

Despite the efforts of \citet{steidelKeckLymanContinuum2018} to produce a clean sample of LyC leakers at \ziii{} by looking for evidence of spectral blending, foreground contamination remains a significant concern for surveys of LyC at high redshift \citep{vanzellaDetectionIonizingRadiation2012,mostardiHIGHRESOLUTIONHUBBLESPACE2015}. A galaxy along the line-of-sight to a \ziii{} source can provide non-ionizing photons that masquerade as rest-frame LyC assuming a single redshift of \ziii{}. As such, \citet{pahlUncontaminatedMeasurementEscaping2021} presented new \textit{HST} measurements of the 15 individual LyC detections in the KLCS sample, which were the objects most likely to be significantly contaminated by foreground light. These data were taken across five survey fields, including seven ACS/F606W (\hstv{}) pointings and 11 WFC3/F125W (\hstj{}) and WFC3/F160W (\hsth{}) pointings. Each pointing was observed for three orbits in each filter. \citet{pahlUncontaminatedMeasurementEscaping2021} also utilized existing \textit{HST} data for one object (Q1549-C25) from \citet{mostardiHIGHRESOLUTIONHUBBLESPACE2015} and \citet{shapleyQ1549C25CLEANSOURCE2016}.
Based on the \vjh{} colors of the subcomponents in the high-resolution, \textit{HST} light profiles, two individual LyC detections were determined to be likely contaminated. An additional 24 LyC non-detections were included in the aforementioned \textit{HST} pointings and were also analyzed. Two of these were found to have associated subcomponents with colors consistent with foreground sources. In total, four galaxies were removed from the KLCS, for a final sample size of 120, including 13 galaxies individually detected in LyC.

\subsection{Photometry and SED fits} \label{sec:sed}

Several galaxy properties can be estimated from broadband photometric measurements. In addition to the $U_nGR$ images used for original photometric selection of \ziii{} candidates \citep[see][]{steidelLymanBreakGalaxies2003}, longer wavelength photometry of the KLCS has been obtained. We summarize the photometric information available for the objects in KLCS in Table \ref{tab:obs}. A subset of these measurements are also summarized and analyzed in \citet{pahlSearchingConnectionIonizingphoton2022}. Specifically, optical, near-IR, and mid-IR data were available for the majority of galaxies in KLCS. We required at least one photometric measurement entirely redward of the Balmer break in order to accurately constrain the stellar populations. The filters that fulfilled this requirement for the KLCS were the $H$, $K_{\rm s}$, and Spitzer/IRAC bands. Thirteen objects objects do not have sufficient IR measurements and were removed from our sample. In addition, we removed two objects with significant scattered light in their ground-based light profiles from nearby objects, and one galaxy identified with multiple redshifts in the original KLCS spectrum.

\begin{table*}
	\centering
	\caption{Photometric bands used in SED modeling.}
	\begin{tabular}{cc}
		\toprule
		Fields   & Photometric bands                                         \\ \midrule
		\midrule
		Q0100 & $U_n^a$, $B^a$, $G^a$, $R_s^a$, $J^b$, $H^b$, $K_s^c$, IRAC1, IRAC2, IRAC3, $H_{\rm 140}$, \hsth{} \\ \midrule
		Q0256 & $U_n^d$, $G^d$, $R_s^d$, $J^c$, $K_s^c$, IRAC1 \\ \midrule
		B20902 & $U_n^{e}$, $G^{d,e}$, $R_s^{d,e}$, $J^f$, $K_s^f$, IRAC1, IRAC2 \\ \midrule
		Q0933    & $U_n^d$, $G^d$, $R_s^d$, $I^d$, $J^c$, $K_s^c$, IRAC1, IRAC2, {\hstv}, {\hstj}, {\hsth}         \\ \midrule
		Q1009 & $U_n^a$, $G^a$, $R_s^a$, $J^g$, $K_s^g$, IRAC1, $H_{\rm 140}$, \hsth{}  \\ \midrule
		Westphal & u*$^h$, g'$^h$, r'$^h$, i'$^h$, z'$^h$, $J^i$, $H^i$, $K_s^{i,f}$, IRAC1, IRAC2, {\hstv}, {\hstj}, {\hsth}        \\ \midrule
		Q1422    & $U_n^e$, $G^e$, $R_s^e$, $K_s^c$, IRAC1, IRAC2, {\hstv}, {\hstj}, {\hsth}            \\ \midrule
		Q1549    & $U_n^a$, $G^a$, $R_s^a$, $J^b$, $H^b$, $K_s^{b,c}$, IRAC1, IRAC2, IRAC3, {\hstv}, {\hstj}, {\hsth}        \\ \midrule
		DSF2237b & $U_n^d$, $G^d$, $R_s^d$, $I^d$, $J^{c,f}$, $K_s^{c,f}$, IRAC1, IRAC2, {\hstv}, {\hstj}, {\hsth}  \\ \bottomrule
	\end{tabular}
	\begin{flushleft}
		$^a$ {Observed with Keck/LRIS.}
		$^b$ {Observed with FourStar at the Magellan Baade 6.5m telescope.}
		$^c$ {Observed with the Multi-Object Spectrometer for Infra-Red Exploration (MOSFIRE) on the Keck I telescope.}
		$^d$ {Observed with the COSMIC prime focus imager on the Palomar 5.08 m telescope \citep[see][]{steidelLymanBreakGalaxies2003}.}
		$^e$ {Observed with the Prime Focus Imager on the William Herschel 4.2m  telescope (WHT) \citep[see][]{steidelLymanBreakGalaxies2003}.}
		$^f$ {Observed with NIRC on the Keck I telescope \citep{shapleyRestFrameOpticalProperties2001}.}
		$^g$ {Observed with the Wide Field Infrared Camera (WIRC) on the Palomar 5.08m telescope.}
		$^h$ {From the Canada-France-Hawaii Telescope (CFHT) Legacy Survey.}
		$^i\;${Observed with CFHT/WIRCam as part of the WIRCam Deep Survey \citep{bielbyWIRCamDeepSurvey2012}.}
	\end{flushleft}
	\label{tab:obs}
\end{table*}

\textit{HST} photometry was included for the 35 objects observed in the \textit{HST} \vjh{} pointings presented in \citet{pahlUncontaminatedMeasurementEscaping2021}. In addition to these 35 objects, 11 objects were covered by at least one \textit{HST} filter, without the full \vjh{} dataset required for contamination analysis. \textit{HST} \hsth{} imaging was also available for four objects in the Q0100 field and two objects in Q1009 \citep{lawHSTWFC3IRMorphological2012a}. We re-measured integrated photometry for all objects in KLCS with \vjh{} \textit{HST} data available largely following the methodology of \citet{pahlUncontaminatedMeasurementEscaping2021} and \citet{pahlSearchingConnectionIonizingphoton2022}. 
In an effort to improve consistency between all photometric measurements, we adopted \textit{HST} magnitudes from the same \textsc{mag\_auto} parameter from  \textsc{sextractor} \citep{bertinSExtractorSoftwareSource1996} that was employed by the ground-based measurements, rather than the isophotal \textit{HST} magnitudes adopted by \citet{pahlUncontaminatedMeasurementEscaping2021} and \citet{pahlSearchingConnectionIonizingphoton2022}. Finally, we note that \textit{HST} $H_{\rm 140}$ measurements are available for three objects in Q0100 and five in Q1009, which were included in our analysis \citep{chenKBSSKCWISurveyConnection2021}.

We attempted to correct the photometry from potential biases resulting from strong emission lines that lie in the bandpass of individual filters. Notably, we used existing Keck/MOSFIRE spectra with coverage of \oii{}, \hb{}, and\oiii{} rest-optical lines to correct broadband $H$ and $K_{\rm s}$ flux measurements, depending on the wavelength of the observed line. We identified eight objects with neither Keck/MOSFIRE spectra nor additional photometry redward of the Balmer break aside from $H$ or $K_{\rm s}$. We removed these objects from the sample to ensure all galaxies in our analysis had at least one trustworthy photometric measurement redward of the Balmer break, free of potential emission-line bias.
In addition, we used \lya{} equivalent widths presented in \citet{steidelKeckLymanContinuum2018} to correct broadband $G$ and \hstv{} flux measurements if the observed wavelength of the line was contained in the respective bandpass. 

\begin{figure*} %
	\centering
	\includegraphics[width=\textwidth]{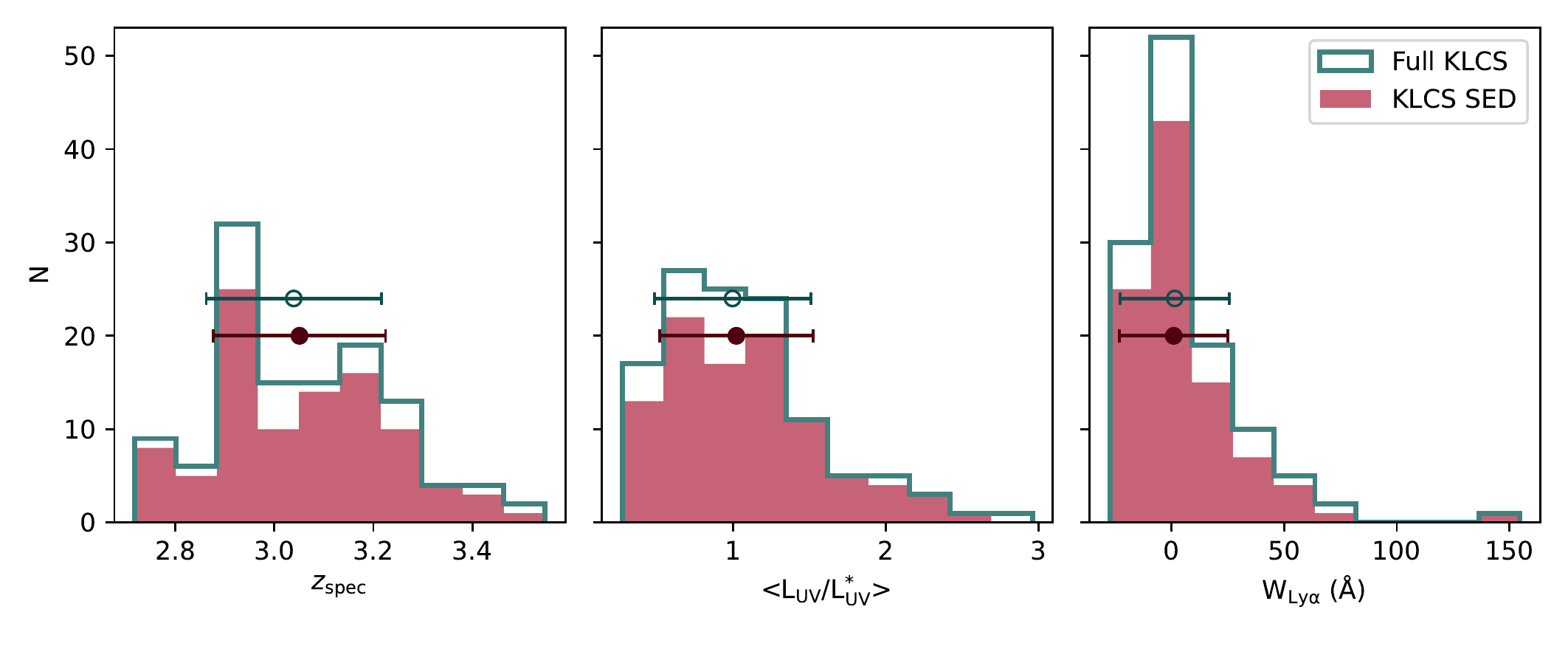}
	\caption{Distributions of $z_{\rm spec}$, \luv{}, and \wlya{} for the KLCS SED sample and the full KLCS. The full KLCS sample includes 120 galaxies from \citet{pahlUncontaminatedMeasurementEscaping2021} while the KLCS SED sample contains the 96 galaxies with photometry appropriate for SED fitting. The median and standard deviation of each distribution are presented respectively as single data points and error bars. The filled circle is the sample median of the KLCS SED sample while the open circle is the sample median of the full KLCS.}
	\label{fig:prop}
\end{figure*}

The final sample with sufficient multi-band photometry for robust stellar population modeling consisted of 96 galaxies, of which 12 were individual LyC detections, which we define as the ``KLCS SED" sample. In Figure \ref{fig:prop}, we display the KLCS SED sample as a function of key observables from \citet{steidelKeckLymanContinuum2018}, including spectroscopic redshift \zspec{}, \lya{} equivalent width \wlya{}, and UV luminosity ($\luv/\luv^*$, where the characteristic luminosity $\luv^*$ corresponds to $M_{\rm UV}^*=-21.0$). We simultaneously present the characteristics of the parent KLCS sample of 120 galaxies. Median \zspec{}, \luv{}, and \wlya{} values for KLCS SED are consistent with those for the full KLCS sample.

In order to estimate stellar-population parameters such as stellar mass (\mstar{}), star-formation rate (SFR), stellar age, and E(B-V) for the galaxies in the KLCS SED sample, we employed SED fits to the multi-band photometry available for these objects. We broadly followed the fitting methodology of \citet{reddyEffectsStellarPopulation2022} \citep[also see][]{pahlSearchingConnectionIonizingphoton2022}. In brief, we utilized BPASS stellar-population synthesis models \citep[BPASS v2.2.1,][]{eldridgeBinaryPopulationSpectral2017} assuming a \citet{chabrierGalacticStellarSubstellar2003} initial-mass function. We assumed a constant star-formation history (SFH) with stellar population ages greater than 50 Myr, such that stellar ages would not be less than typical dynamical-timescales of \ziii{} star-forming galaxies \citep{reddyCharacteristicStarFormation2012}. We adopted constant SFHs as they have been shown to reproduce independent measurements of SFR for galaxies $z\geq1.5$ \citep{reddyCharacteristicStarFormation2012}. Constant SFHs may also provide a better description of galaxies at the stellar masses of our sample ($\sim10^9-10^{10.5}\:\msun$), which may have less bursty SFHs than galaxies at lower masses \citep[e.g.][]{dominguezConsequencesBurstyStar2015}.
We adopted assumptions of metallicity of 0.14 times solar and an SMC dust attenuation curve \citep{gordonQuantitativeComparisonSmall2003}. We examined each SED fit individually for outlier photometric measurements, and dropped Spitzer/IRAC data with clear evidence of blending from nearby sources.

\subsection{Binning strategy and spectral modeling} \label{sec:spec}

While $f_{900}$ can be measured for each object individually, constraining the LyC leaking in the vicinity of a galaxy requires an understanding of the attenuation of the signal from neutral hydrogen along the line of sight in the IGM and CGM. The transmission of LyC emission varies significantly from sightline to sightline at the redshifts of our sample, introducing large uncertainties on individual LyC measurements \citep{steidelKeckLymanContinuum2018}. To circumvent this sightline to sightline variability, we used binned subsamples and composite spectra that reflect average effects of IGM and CGM attenuation on the LyC spectral region as in \citet{steidelKeckLymanContinuum2018}. In order to understand how ionizing-spectral properties vary with the properties produced by SED fits described in the previous section, we binned the KLCS SED sample as a function of \mstar{}, SFR, E(B-V), age, and specific star-formation rate (sSFR; sSFR$\equiv$SFR/\mstar{}). We created three bins for each property, each containing 32 galaxies, to ensure that the mean IGM+CGM transmission is known with $\lessapprox10\%$ uncertainty \citep{steidelKeckLymanContinuum2018} in subsequent composite spectra. We also created an ``all" sample, containing all 96 galaxies from KLCS SED, and binned subsamples for \wlya{} and \luv{}.

For each subsample, we generated composite spectra representing the average spectral properties of the component galaxies. Following the methodology of \citet{steidelKeckLymanContinuum2018} \citep[also see][]{pahlUncontaminatedMeasurementEscaping2021,pahlSearchingConnectionIonizingphoton2022}, each individual spectrum is first normalized to the average flux density in the non-ionizing UV spectral region, $1475-1525$\AA{} in the rest frame. Using the set of normalized spectra for each binned sample, we then computed the sigma-clipped mean of the distribution of flux densities at each rest-frame wavelength increment, with clipping applied at 3$\sigma$. We did not apply sigma clipping to the Ly$\alpha$ spectral region ($1200-1230$\AA{}) in order to conserve the inferred composite Ly$\alpha$ profile. The error on the mean flux density at each wavelength was propagated from the values of individual error spectra.

\begin{figure*} %
	\centering
	\includegraphics[width=\textwidth]{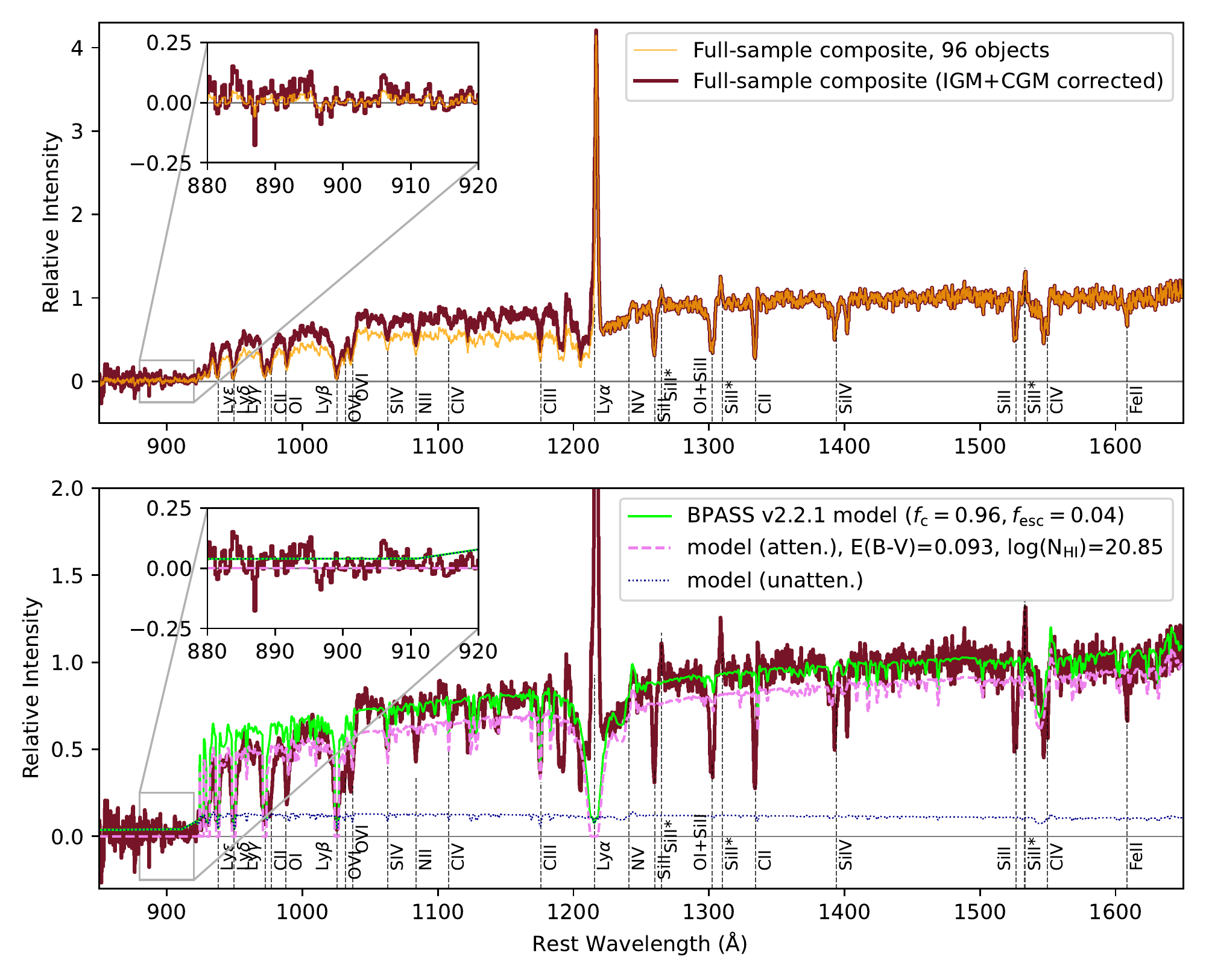}
	\caption{Composite spectrum for the ``all" (full KLCS SED, 96 galaxy) sample.
		\textbf{Top:} The ``all" composite alongside the same spectrum corrected from the average attenuation from the IGM and CGM at the mean redshift of the composite, $z_{\rm mean}=3.05$. The uncorrected spectrum is shown with a thin, orange curve, while the corrected composite is shown with a thick, maroon curve. An inset is included to highlight the LyC spectral region.
		\textbf{Bottom:} IGM- and CGM-corrected composite spectrum alongside the best-fit spectrum from the modeling process. The corrected composite is again shown with the thick, maroon curve. The best-fit BPASS model is presented as a thin, green curve. This model is summed from two component spectra, an attenuated portion displayed as a dashed, pink line, and an unattenuated portion displayed as a dotted, blue line, as per the ``holes" model of \citet{steidelKeckLymanContinuum2018}. An inset is included to highlight the LyC spectral region. The free parameters values of the fit that produced the model curves are $f_{\rm c}=0.96$, E(B-V)$=0.093$, and log(N$_{\rm HI}$/cm$^{-2}$)$=20.85$.
	}
	\label{fig:spec}
\end{figure*}

For each composite spectrum, we computed \fobs{}, which is the ratio between the average flux densities in the LyC region ($880-910$\AA{}, $f_{900}$) and the non-ionizing UV continuum ($1475-1525$\AA{}, $f_{1500}$). While this ratio is useful for discerning the average observed ionizing photon leakage relative to the non-ionizing ultraviolet luminosity density, as discussed above, the spectra must be corrected for lowered transmission from the IGM in the LyC region in order to understand the average effect of LyC leakage has on its environment. We corrected the spectra using average ``IGM+CGM" transmission functions from \citet{steidelKeckLymanContinuum2018}, calculated at the mean redshift of each composite subsample, and based on the statistics of H~\textsc{i} absorption systems along QSO sightlines presented by \citet{rudieGaseousEnvironmentHighz2012} and \citet{rudieColumnDensityDistribution2013}. 
To demonstrate the characteristics of the composite spectra used in our analysis, we display the ``all" composite before and after the IGM+CGM transmission correction in the upper panel of Figure \ref{fig:spec}. Using corrected spectra, we repeated the measurement of the ratio of $f_{900}$ to $f_{1500}$, defined as \fout{}, which applies to the ratio that would be observed at $50\:$proper kpc from galaxy center \citep[see][]{steidelKeckLymanContinuum2018}.

While \fout{} is a useful empirical measurement of leaking LyC, \fesc{} remains extensively used in reionization modeling. In order to calculate the average \fesc{} for each subsample, we require both an understanding of the intrinsic UV spectrum of the galaxies and the average effects from any intervening gas in the ISM. Thus, \fesc{} is dependent on the assumed stellar population synthesis model, and we follow the well-motivated choices for such models discussed in \citet{steidelKeckLymanContinuum2018}.
We introduced consistency between our multi-wavelength and spectroscopic modeling by again using the BPASS stellar-population synthesis models of \citet{eldridgeBinaryPopulationSpectral2017}. We coupled these models with an SMC extinction curve \citep{gordonQuantitativeComparisonSmall2003} and a range of E(B-V) from 0.0 to 0.6, and assumptions of metallicity of 0.07 times solar.
This metallicity is similar to that assumed for the SED fitting and is consistent with the spectral modeling of \citet{steidelKeckLymanContinuum2018} and \citet{pahlUncontaminatedMeasurementEscaping2021}.
We model the ISM geometrically using the ``holes" approach, which assumes LyC light escapes through a patchy neutral-phase gas \citep{zackrissonSpectralEvolutionFirst2013,reddyConnectionReddeningGas2016,reddyEffectsStellarPopulation2022}. The free parameters of the fit included the neutral gas covering fraction $f_{\rm c}$, the column density of neutral hydrogen N$_{\rm HI}$, and the dust attenuation from the foreground gas E(B-V)$_{\rm cov}$ (i.e., the uncorrected portion is assumed to be dust free).	In general, \fesc{} is defined from $f_{\rm c}$, where $\fesc=1-f_{c}$. To demonstrate the fitting process, we display a fit to the corrected full-sample composite in the lower panel of Figure \ref{fig:spec}. Here, the modeled spectrum in green is split into an unattenuated (pink) and attenuated (blue) portion, representing the light that either escaped through clear sightlines in the ISM or was partially reduced by intervening material, respectively.

In order to estimate the uncertainty in average escape parameters for a given set of galaxies, we must understand the level of variability induced from sample construction. With the goal of understanding how sample variance affects generated composite spectra, we used bootstrap resampling of the galaxies of each binned subsample. We generated 100 sets of 32 galaxies (the number of galaxies in each bin) by performing draws from a given subsample with replacement. We subsequently created composite spectra and measured ionizing-photon escape for each random draw, using the process described earlier in this section. The mean and standard deviation of the \fobs{}, \fout{}, and \fesc{} distributions generated from the 100 composite spectra were used as the fiducial value and error estimate for the corresponding binned sample. The errors determined from this bootstrap resampling were larger than those associated with measurement uncertainty and average IGM+CGM transmission variability.

\section{Results} \label{sec:res}

\begin{figure*} %
	\centering
	\includegraphics[width=\textwidth]{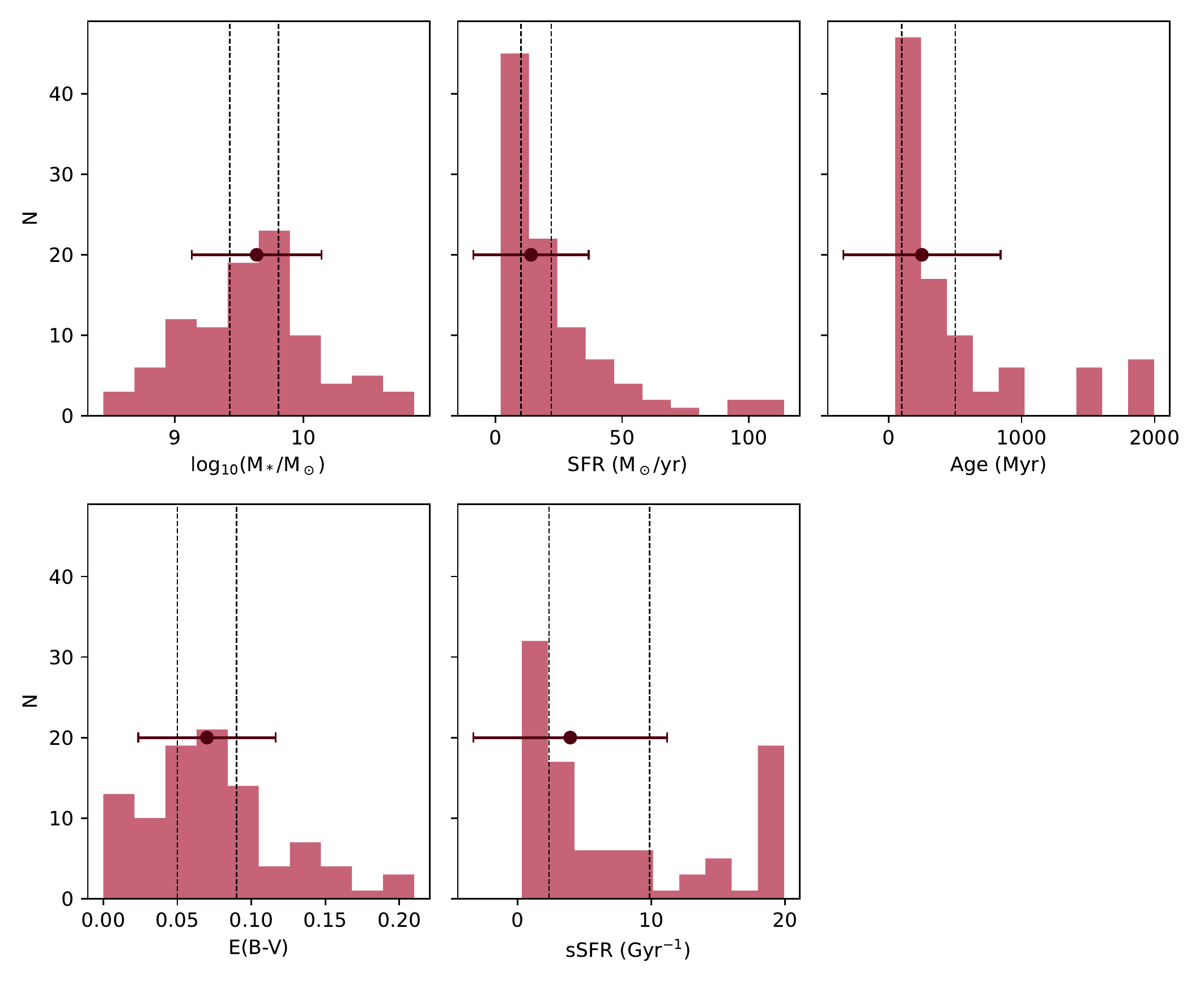}
	\caption{
		Galaxy property distributions of the KLCS SED sample. The five properties displayed here were inferred from SED fits performed for each galaxy. The median and standard deviations with respect to a given measurement are presented as single data points with error bars. The edges of bins used for generation of composite spectra are shown as vertical, dashed lines. The full sample was sorted according to each galaxy property and divided into three equal-sized bins (n=32), which were then used to generate composite spectra.
	}
	\label{fig:sedprop}
\end{figure*}

Based on SED fits, we estimated \mstar{}, SFR, stellar age, E(B-V) and sSFR for each galaxy. We present the distribution of these SED-modeled parameters for the KLCS SED sample in Figure \ref{fig:sedprop}. The median and standard deviation of each respective measurement distribution are displayed, respectively, as dark points and error bars. In the figure, 
we use dashed, vertical lines to indicate the edges of the three equal-sized (n=32) samples that comprise the bins for generating composite spectra.

\begin{figure*} %
	\centering
	\includegraphics[width=\textwidth]{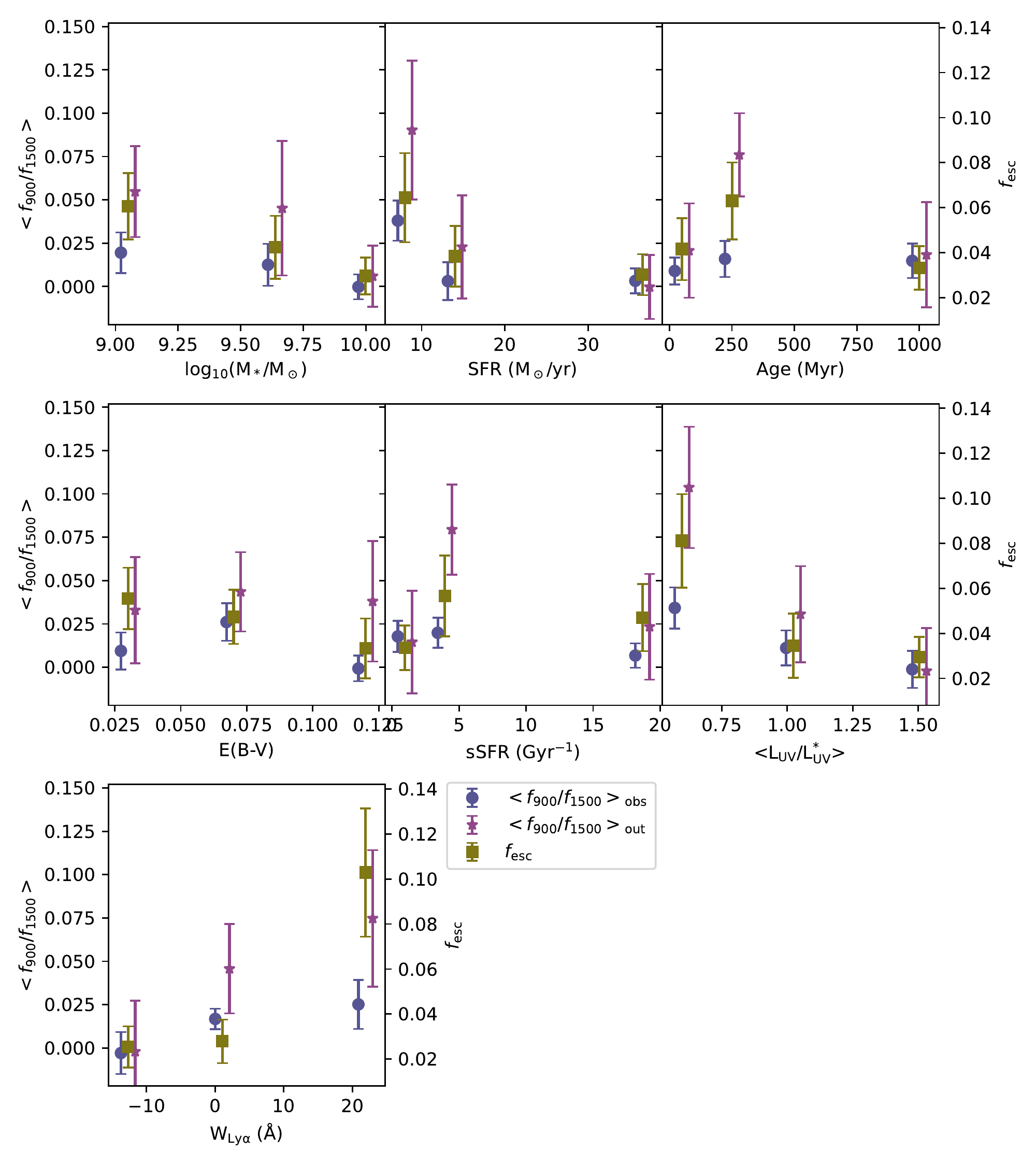}
	\caption{Different measures of ionizing escape for the KLCS SED sample binned according to several galaxy properties. Each escape parameter is presented as a function of the median of the respective galaxy property of the binned subsample. Measurements of \fobs{} are presented as blue circles and are based on uncorrected, composite spectra. Measurements of \fout{} are presented as purple stars and are based on composite spectra corrected for attenuation of the IGM+CGM; estimates of \fesc{} are presented as yellow squares and are based on SPS model fits and modeling of the ISM according to the ``holes" model \citep{steidelKeckLymanContinuum2018,reddyConnectionReddeningGas2016,reddyEffectsStellarPopulation2022}. Blue circles and purple stars are shifted left and right, respectively, for visual clarity.
	}
	\label{fig:alllyc}
\end{figure*}

A composite spectrum was generated for each binned sample detailed in Figure \ref{fig:sedprop}, and three estimates of ionizing-photon escape were measured, as described in Section \ref{sec:spec}. The first, \fobs{}, is the ratio of ionizing to non-ionizing flux density directly observed in the composite generated from individual spectra. The second, \fout{}, is the same ratio instead measured from a composite corrected for mean line-of-sight attenuation from the IGM and CGM. Finally, \fesc{} is a parameter estimated via stellar-population synthesis and ISM modeling of the full rest-UV composite. We display the three measurements of ionizing escape and their respective errors for each subsample binned as a function of galaxy property in Figure \ref{fig:alllyc}.

To determine whether ionizing-photon escape is correlated with measured galaxy properties, we define a ``significant" correlation as fulfilling two criteria: the escape parameter varies monotonically across the three bins, and the difference between the escape parameter in the highest and lowest bins was $>1\sigma$. Using \fesc{} as an example, we define a $>1\sigma$ difference as 
\begin{equation}
|f_{\rm esc,highest}-f_{\rm esc,lowest}|>\sqrt{(\sigma_{f_{\rm esc,highest}})^2 + (\sigma_{f_{\rm esc,lowest}})^2},
\end{equation}
where $f_{\rm esc,highest}$ is measured from the third tertile of a given galaxy property, $f_{\rm esc,lowest}$ is from the first tertile, and $\sigma_{f_{\rm esc,highest}}$ and $\sigma_{f_{\rm esc,lowest}}$ are their corresponding errors derived as described in Section \ref{sec:spec}. 

As seen in the upper left panel of Figure \ref{fig:alllyc}, both \fout{} and \fesc{} are significantly, negatively correlated with \mstar{} in the KLCS SED sample, such that lower-mass galaxies have higher escape fractions. The two measures of LyC escape are also significantly, negatively correlated with SFR, shown in the upper central panel. While we find no significant correlation with \fout{} and E(B-V), we do find a significant, negative correlation with \fesc{} and E(B-V), as displayed in the upper right panel. Discrepancies between these two parameters of LyC escape arise from the fact that the fitting process to determine \fesc{} incorporates additional information from the composite spectrum, including the Lyman series absorption features and the UV spectral shape. Correlation between ionizing-escape parameters and E(B-V) is expected in our analysis considering that neutral gas and dust are spatially associated \citep{reddyConnectionReddeningGas2016,duRedshiftEvolutionRestUV2018,pahlRedshiftEvolutionRestUV2020}. We also see an anti-correlation between \fesc{} and E(B-V) as inferred from spectral modeling. Finally, we find no significant correlation between \fout{} or \fesc{} and stellar age or sSFR, seen in the middle left and center panels of Figure \ref{fig:alllyc}, respectively. 

We also note that \fesc{} and \fout{} are significantly correlated with \wlya{} and \luv{} in this analysis, seen in the middle right and bottom panels of Figure \ref{fig:alllyc}, mirroring the results of the full KLCS presented in \citet{steidelKeckLymanContinuum2018} and \citet{pahlUncontaminatedMeasurementEscaping2021}. The positive trend between \fesc{} and \wlya{} has also been confirmed in additional $z\sim3-4$ LyC surveys \citep{marchiNewConstraintsAverage2017,marchiLyaLymanContinuumConnection2018,fletcherLYMANCONTINUUMESCAPE2019,begleyVANDELSSurveyMeasurement2022}. In \citet{pahlSearchingConnectionIonizingphoton2022}, we argued that recovering these well-established spectral trends is important for determining whether a sample is sufficiently large and representative for examining relationships between \fesc{} and other galaxy properties. 
Considering the KLCS SED sample has both the size (n=96) and dynamic range of galaxy properties to confidently recover trends between \fesc{} and \wlya{}/\luv{}, we conclude that the KLCS SED sample is sufficient and representative, fulfilling the requirements for determining the trends between \fesc{} and galaxy property presented in this section.

\section{Discussion} \label{sec:disc}

The connections between \fesc{} and galaxy properties at \ziii{} provide key insights into the physics of ionizing-photon escape, and also indicate the most appropriate assumptions for \fesc{} at even higher redshift, during the epoch of reionization. We find that galaxies with higher \fesc{} tend to have lower E(B-V), which is consistent with a physical picture in which dust is spatially coincident with neutral-phase gas in a galaxy, such that an ISM with a higher neutral-gas covering fraction will be both dustier and have lower associated \fesc{} \citep{reddyConnectionReddeningGas2016,duRedshiftEvolutionRestUV2018,pahlRedshiftEvolutionRestUV2020}. In addition, ionizing photons are more attenuated by dust than non-ionizing photons \citep{reddySPECTROSCOPICMEASUREMENTSFARULTRAVIOLET2016}.
We find a negative trend between both \fesc{} and \fout{} and \mstar{}, highlighting the fact that more massive galaxies at \ziii{} have conditions that are less conducive to LyC escape. This relationship is likely due to the fact that more massive galaxies tend to be dustier \citep[e.g.,][]{whitakerConstantAverageRelationship2017,mclureDustAttenuationStarforming2018}. Finally, the lack of trend between \fesc{} and either stellar age or sSFR is in tension with the physical picture advanced in simulations that bursts of recent star formation induce favorable channels in the ISM and CGM for ionizing photons to escape \citep{maNoMissingPhotons2020}. In this section, we introduce comparisons between our results and recent LyC surveys both at \ziii{} and in the local Universe. We also connect our ionizing-photon escape trends or lack thereof with radiative transfer modeling of simulated galaxies and the predictions from reionization models.

\subsection{Comparison to related observational surveys} \label{sec:discobs}

\begin{figure} %
	\centering
	\includegraphics[width=\columnwidth]{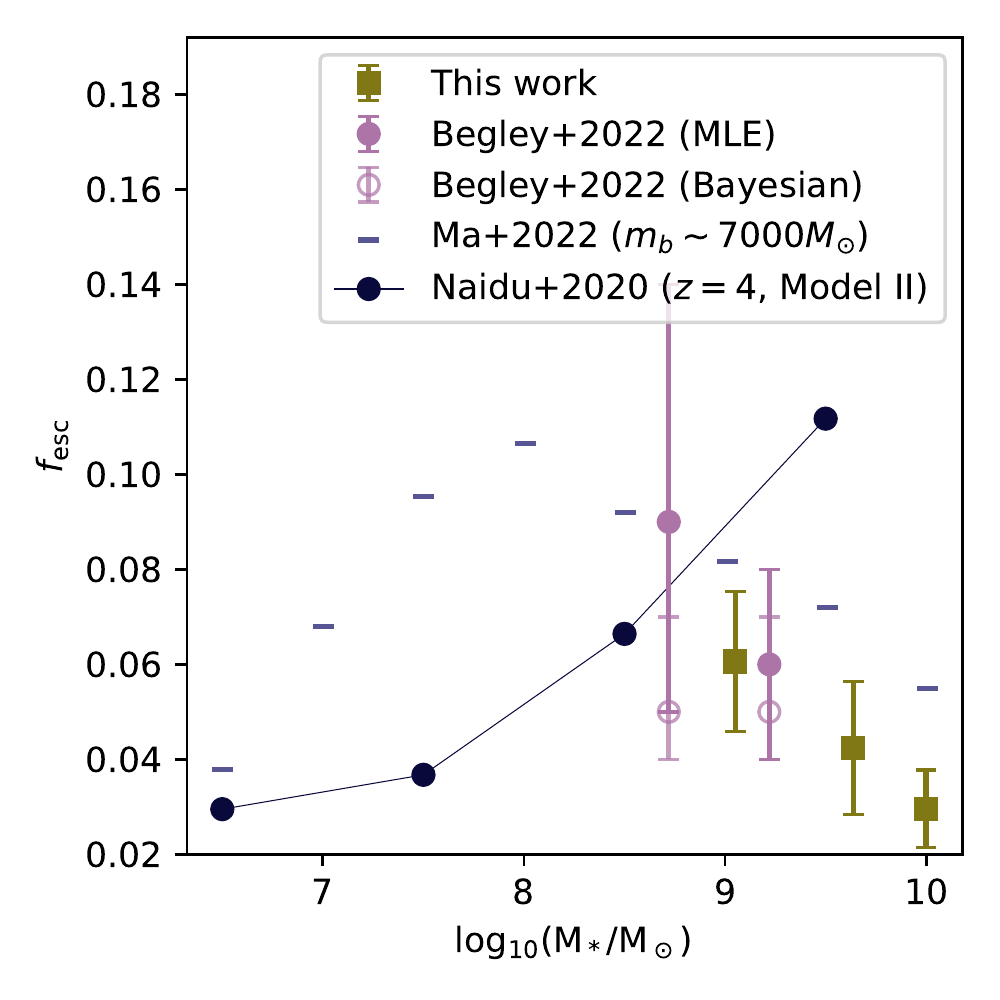}
	\caption{Inferred \fesc{} as a function of stellar mass from this work alongside trends from observation and modeling. Estimates of \fesc{} for three bins of increasing \mstar{} for the KLCS SED sample are presented as yellow boxes, and are identical to values presented in Figure \ref{fig:alllyc}. The \fesc{} constraints from two bins of increasing \mstar{} from 148 $z\sim3.5$ galaxies from VANDELS \citep{begleyVANDELSSurveyMeasurement2022} are displayed as purple circles. Solid, purple circles represent \fesc{} fit by maximum-likelihood analysis, while skeletal, purple circles represent \fesc{} fit by Bayesian analysis. Predictions for \fesc{} from the FIRE-2 cosmological simulations at a particle mass of $m_B\sim7000\msun$ are displayed as horizontal bars \citep{maNoMissingPhotons2020}. Predicted \fesc{} as a function of \mstar{} at $z\sim4$ for the fiducial model of \citet{naiduRapidReionizationOligarchs2020} are displayed as dark circles.
	}
	\label{fig:lycmass}
\end{figure}

Direct comparisons can be made between our reported trends between \fesc{} and galaxy property and those found in recent LyC surveys at \ziii{}. Of particular note are the recent photometric LyC measurements of 148 galaxies from the VANDELS survey at $3.35\leq z\leq3.95$ \citep{begleyVANDELSSurveyMeasurement2022}. These authors constrained the average \fesc{} of the sample as $\langle \fesc \rangle=0.07\pm0.02$, consistent with $\langle \fesc \rangle=0.06\pm0.01$ measured from the uncontaminated KLCS \citep{pahlUncontaminatedMeasurementEscaping2021}. The VANDELS LyC sample was binned in two as a function of a variety of galaxy properties. A positive correlation between \fesc{} and \wlya{} and a negative correlation between \fesc{} and \luv{} reported in the VANDELS analysis aligns with the correlations found in the KLCS SED sample and the full KLCS \citep{steidelKeckLymanContinuum2018,pahlUncontaminatedMeasurementEscaping2021}. Best-fit \fesc{} values were also calculated for two bins of increasing \mstar{} in the VANDELS sample, which we display alongside our \fesc{} vs. \mstar{} measurements for the KLCS SED sample in Figure \ref{fig:lycmass}. A weak anti-correlation was observed between \fesc{} and \mstar{} in the VANDELS analysis when utilizing maximum-likelihood estimation (MLE) to determine \fesc{}. No correlation was found when using a Bayesian estimate of \fesc{}. The trend between the VANDELS MLE \fesc{} values and \mstar{} is remarkably consistent with our results. In addition, the reported \fesc{} values for both Bayesian and MLE methods from VANDELS are consistent with our \fesc{} constraints at comparable \mstar{}, however we note that the VANDELS results use modeling that more closely resembles the ``screen" model of \citet{steidelKeckLymanContinuum2018}, rather than the ``holes" model used in this work. Using the ``screen" model results in $\sim30\%$ higher $\langle \fesc \rangle$ than using the ``holes" model in the KLCS \citep{steidelKeckLymanContinuum2018}, which is still consistent with the VANDELS results. The VANDELS analysis also recovered a significant anti-correlation between \fesc{} and UV dust attenuation, where dust attenuation was quantified in terms of the UV slope, $\beta$. These results are qualitatively consistent with the anti-correlation between \fesc{} and E(B-V) we present in the central left panel of Figure \ref{fig:alllyc}.

Stellar population parameters have also been explicitly correlated with \wlya{} in galaxy surveys at $z\sim2-5$. These trends are informative for interpreting LyC escape considering that the strength of Ly$\alpha$ emission is similarly modulated by the neutral gas covering fraction \citep[e.g.,][]{steidelStructureKinematicsCircumgalactic2010,steidelDiffuseLyaEmitting2011,steidelKeckLymanContinuum2018,verhammeUsingLymanDetect2015,reddyConnectionReddeningGas2016}. Stacks of rest-UV spectra at $z\sim2-5$ have demonstrated anti-correlations between \wlya{} and both \mstar{} and SFR \citep{duRedshiftEvolutionRestUV2018,pahlRedshiftEvolutionRestUV2020}, mimicking the anti-correlations between \fesc{} and these parameters that we presented in Figure \ref{fig:alllyc}. Meanwhile, surveys at this redshift have shown either no strong correlation between \wlya{} and age \citep{duRedshiftEvolutionRestUV2018,pahlRedshiftEvolutionRestUV2020} or only a weak correlation \citep{reddyEffectsStellarPopulation2022}. These analyses of Ly$\alpha$ escape in combination with our \fesc{} trends indicate that stellar age may not as closely linked to the configuration of neutral-phase gas in the ISM and CGM of a galaxy as much as other galaxy properties, such as \mstar{}, \luv{}, E(B-V), and SFR.

While the $z\sim3-4$ universe is an excellent laboratory to test LyC escape physics in galactic environments more similar to those at $z>6$, LyC surveys in the local universe are afforded advantages such as the ability to examine the direct ionizing signals from intrinsically fainter galaxies in the dwarf galaxy regime, which may dominate the ionizing background during the epoch of reionization \citep{robertsonCosmicReionizationEarly2015,finkelsteinConditionsReionizingUniverse2019}. In addition, local surveys avoid the sightline variability of the IGM that necessitates binning at \ziii{} \citep{steidelKeckLymanContinuum2018}, enabling constraints on \fesc{} for individual objects. 

The Low-redshift Lyman Continuum Survey (LzLCS) analyzed 66 galaxies at $z=0.2-0.4$ observed with the \textit{HST}/COS, and reported 35 galaxies individually detected in LyC \citep{fluryLowredshiftLymanContinuum2022,fluryLowredshiftLymanContinuum2022a}. The galaxies were indirectly selected to be strongly leaking using [O~\textsc{iii}]\ensuremath{\lambda5007}/\oii{}, SFR surface density, and UV spectral slope, in contrast the LBG-selected KLCS. The correlation between a number of galaxy properties and \fesc{} were considered, where \fesc{} was inferred from stellar-population synthesis fits to COS UV spectra, similar to our determinations of \fesc{} for the KLCS SED sample. 
The LzLCS \fesc{} values appear to decrease as a function of increasing \mstar{}, consistent with the negative trend we present in the upper left panel of Figure \ref{fig:alllyc}. However, the correlation coefficient between \fesc{} and \mstar{} was determined not to be significant, mirroring other local explorations of the two variables \citep{izotovLymanContinuumLeakage2021}. 
Augmenting this result, an examination of the LzLCS sample in tandem with archival observations (totaling 89 star-forming galaxies at $z\sim0.3$) found that galaxies at lower \mstar{} tend to have both bluer spectral slopes and higher \fesc{} \citep{chisholmFarUltravioletContinuumSlope2022}. This analysis focused on the strong inverse correlation found between \fesc{} and the UV spectral slope at 1550\AA{} ($\beta_{1500}$) in the expanded sample. While UV spectral slope can encapsulate both the intrinsic spectral slope of the stellar population and the degree of dust reddening in the UV, $\beta_{1500}$ was strongly correlated with E(B-V) and uncorrelated with stellar age, indicating that $\beta_{1500}$ for the LzLCS is primarily reflecting the degree of dust reddening. Apparent anti-correlations between \fesc{} and $\beta_{1500}$ are supported by our anti-correlation of \fesc{} and E(B-V) seen in the center left panel of Figure \ref{fig:alllyc}. Nonetheless, \citet{chisholmFarUltravioletContinuumSlope2022} make predictions for \fesc{} vs. $M_{\rm UV}$ at $z\sim3$ that are too low when compared to the $\langle \fesc \rangle=0.06\pm0.01$ of the KLCS (at $M_{\rm UV}\sim-21$), despite reproducing our qualitative relationship between \fesc{} and \luv{} shown in the center right panel of Figure \ref{fig:alllyc}.

A weaker, but still significant trend of \fesc{} and sSFR was also observed in the LzLCS, contrasting with the lack of trend between \fesc{} and sSFR that we presented in the central panel of Figure \ref{fig:alllyc}. We note that the LzLCS modeling allows for arbitrarily-short stellar ages, in contrast with our SED fitting procedure, which ensures ages are greater than typical dynamical timescales ($50\:$Myr).
Finally, \fesc{} is strongly correlated with \wlya{} in the LzLCS analysis, which is broadly consistent with the correlation found for both the KLCS SED sample in Figure \ref{fig:alllyc} and the full KLCS in \citet{steidelKeckLymanContinuum2018} and \citet{pahlUncontaminatedMeasurementEscaping2021}.

We also note that strong correlations are found between \fesc{} and Ly$\alpha$ peak separation ($v_{\rm sep}$) and star-formation rate surface density (\sfrd{}) in the LzLCS. These potential tracers of \fesc{} remain unconfirmed at \ziii{}, and, in particular, the number of galaxies with \textit{HST} imaging in the KLCS remains insufficient for testing the connection between \fesc{} and \sfrd{} \citep{pahlSearchingConnectionIonizingphoton2022}. However, future work will examine potential connections between \fesc{} and $v_{\rm sep}$ at \ziii{} by leveraging higher-resolution spectroscopy of the Ly$\alpha$ profiles of KLCS galaxies and \fesc{} and \sfrd{} in the VANDELS survey.

\subsection{Comparison to models}

Theoretical predictions for \fesc{} in a variety of galactic environments can help elucidate fundamental relationships between galactic physics and escaping ionizing radiation in the earliest galaxy populations, where direct LyC detections are impossible. The Feedback in Realistic Environments \citep[FIRE-2;][]{hopkinsFIRE2SimulationsPhysics2018} project was coupled with radiative transfer in post-processing to examine \fesc{} in cosmological zoom-in simulations of galaxies, evolved down to $z=5$ \citep{maNoMissingPhotons2020}. An increase of \fesc{} with increasing mass was found up to log(\mstar{}/\msun{})$\sim8$, and a subsequent decrease in \fesc{} was found at log(\mstar{}/\msun{})$>8$. The increasing relationship between \fesc{} and \mstar{} was determined to be due to an increasing efficiency of star formation and feedback, while the decrease at the high-mass end can be explained by increasing dust attenuation. We display this trend as dark, horizontal bars in Figure \ref{fig:lycmass}, specifying the simulation resolution that extends to the stellar masses of our sample (baryonic particle mass $m_b\sim7000M_{\odot}$). The negative trend between \fesc{} and \mstar{} at log(\mstar{}/\msun{})$>8$ found in \citet{maNoMissingPhotons2020} is consistent with the negative trend we find in the KLCS SED sample, which has a median log(\mstar{}/\msun{})$=9.6$. We do find overall lower \fesc{} values than the FIRE-2 results at fixed \mstar{}. This discrepancy may be expected considering higher \fesc{} values were found with increasing redshift at fixed \mstar{} in the simulated galaxies.

Evidence of a turnover in the relationship between \fesc{} and \mstar{} was also found in \citet{kostyukIonizingPhotonProduction2022}, which utilized the IllustrisTNG \citep{marinacciFirstResultsIllustrisTNG2018,naimanFirstResultsIllustrisTNG2018,nelsonFirstResultsIllustrisTNG2018,nelsonFirstResultsTNG502019,pillepichFirstResultsIllustrisTNG2018,springelFirstResultsIllustrisTNG2018} cosmological simulations coupled with the radiative transfer code CRASH \citep{grazianiCRASH3CosmologicalRadiative2013}. These authors also found significant scatter in the relationship between \fesc{} and \mstar{}, due to both differences in ionizing photon production rates and the distribution of stars within the neutral ISM.

Finally, \citet{maNoMissingPhotons2020} also explored potential synchronization of periods of intense star formation and elevated \fesc{} values. They find that feedback from star formation clears sightlines in the ISM and CGM of a galaxy, creating favorable conditions for ionizing photons to escape. This process leads to a correlation between a burst of star-formation and high \fesc{}, albeit with a few Myr time delay. If true, one might expect a higher \fesc{} in galaxies with shorter stellar ages and elevated sSFR. We find no correlation between \fesc{} and these two properties in the KLCS SED analysis, as shown in Figure \ref{fig:alllyc}. The absence of an observed trend could be explained by a less bursty SFH than those found in \citet{maNoMissingPhotons2020}, which would reduce potential dependencies between \fesc{} and stellar age.

\subsection{Implications for reionization}

Models of reionization and their predicted timelines are built upon assumptions regarding \fesc{}, which are impossible to constrain directly in the reionization era. Some assume single values of \fesc{} for all galaxies for simplicity  \citep[typically $10-20\%$;][]{robertsonCosmicReionizationEarly2015,ishigakiFulldataResultsHubble2018}, others assume that \fesc{} depends on halo mass \citep{finkelsteinConditionsReionizingUniverse2019}, or that \fesc{} depends on one particular galaxy property \citep{naiduRapidReionizationOligarchs2020,mattheeReSolvingReionization2022}.

We compare our \fesc{} vs. \mstar{} trend to the predictions of the fiducial model of \citet{naiduRapidReionizationOligarchs2020}, which concludes that reionization is ``oligarchical," such that the most luminous, massive ($M_{\rm UV}<-18$ and log(\mstar{}/\msun{})$>8$) galaxies at $z>6$ contribute the bulk of the ionizing photon budget. In this model, a direct relationship between \fesc{} and \sfrd{} is assumed such that \fesc{} = 1.6$\times\sfrd^{0.4}$. As massive and UV bright galaxies tend to have high \sfrd{}, the assumed connection between \fesc{} and \sfrd{} results in a positive relationship between \fesc{} and \mstar{}. We display this trend determined at $z=4$ as dark circles in Figure \ref{fig:lycmass}. The trend we observe between \fesc{} and \mstar{} at \ziii{} is inconsistent with the direction and magnitude of the model curve within the mass range where the model and observations overlap. Specifically, we show that \fesc{} decreases with increasing \mstar{} at log(\mstar/\msun)$\sim9.5$. We also find significantly lower $\langle \fesc \rangle$ at fixed \mstar{} than is predicted by the model. Some of this offset at fixed mass is likely due to the average value of $\langle \fesc \rangle=0.09\pm0.01$ from \citet{steidelKeckLymanContinuum2018}, used as a constraint in the model, considering $\langle \fesc \rangle$ of the KLCS was corrected to $\langle \fesc \rangle=0.06\pm0.01$ after removal of foreground contamination \citep{pahlUncontaminatedMeasurementEscaping2021}. Additionally, the fiducial model of \citet{naiduRapidReionizationOligarchs2020} does not explicitly consider dust, which we find is a significant factor modulating the escape fraction of galaxies in our sample. To conservatively match observed relationships between \fesc{} and \mstar{} at $z\sim3$, assumed \fesc{} values of galaxies at $\mstar>10^9\msun$ should be no higher than the \fesc{} values of $\mstar\sim10^{8.5}\msun$ galaxies.
Specifically, the \fesc{} value for the most massive \mstar{} datapoint from \citet{naiduRapidReionizationOligarchs2020} should shift to become lower than or equal to the \fesc{} value for the second-most-massive datapoint.
Meeting this requirement at $z\sim4$, which is the closest point of contact between the \citet{naiduRapidReionizationOligarchs2020} model and our observations, would require a reduction in \fesc{} for $\mstar>10^9$ galaxies by a factor of two  in the fiducial model of \citet{naiduRapidReionizationOligarchs2020}. Satisfying this criterion at $z>6$ during the epoch of reionization would require a similar reduction. This adjustment would significantly shift the burden of reionization to lower mass galaxies.

The rapidity of reionization depends strongly on the population of galaxies that dominates the ionizing emissivity over cosmic time. Our results indicate that fainter, less massive galaxies with lower dust content have conditions favorable for escaping ionizing radiation, broadly consistent with other recent LyC observations at $z\sim3$ \citep{begleyVANDELSSurveyMeasurement2022} and in the local Universe \citep{fluryLowredshiftLymanContinuum2022,fluryLowredshiftLymanContinuum2022a,chisholmFarUltravioletContinuumSlope2022}. If these trends were present within the epoch of reionization, the process of reionization may have started early and progressed gradually, such that the IGM neutral fraction is $20\%$ at $z\sim7$ \citep{finkelsteinConditionsReionizingUniverse2019}, in slight tension with neutral fraction constraints from Ly$\alpha$ damping wing measurements \citep{boltonHowNeutralIntergalactic2011,greigAreWeWitnessing2017,banados800millionsolarmassBlackHole2018}. 
\citet{chisholmFarUltravioletContinuumSlope2022} calculate the ionizing emissivity between $z\sim4-8$ using empirical relations between \fesc{} and $\beta_{1500}$ found in the LzLCS, which are consistent with our results connecting \fesc{} and E(B-V), and match constraints indicating that the ionizing emissivity flattens out at $z<5.5$ \citep{beckerNewMeasurementsIonizing2013,beckerMeanFreePath2021}. However, as noted in Section \ref{sec:discobs}, average \fesc{} values assumed by \citet{chisholmFarUltravioletContinuumSlope2022} are too low at $z\sim3$ when compared to the KLCS. 
As an alternative scenario, \citet{mattheeReSolvingReionization2022} instead directly tie \fesc{} to the strength of Ly$\alpha$ emission and build a model that produces rapid reionization and a flattened evolution of the ionizing emissivity at $z<6$. 
Predicted trends between \fesc{} and \luv{} from the model of \citet{mattheeReSolvingReionization2022} qualitatively match the KLCS anti-correlation, but underpredict the average \fesc{} at \ziii{}. 
Nonetheless, such a prescription is promising considering that the relationship between \fesc{} and \wlya{} appears be one of the most fundamental in our analysis.
The \citet{mattheeReSolvingReionization2022} model assumes $\fesc=50\%$ for half of Ly$\alpha$ emitters with $L_{\rm Ly\alpha}>10^{42}\textrm{erg s$^{-1}$}$, based on the Ly$\alpha$ line profile shapes of $z\sim2$ Ly$\alpha$ emitters \citep{naiduSynchronyProductionEscape2022}. Our ongoing spectroscopic observing program to explore the connection between Ly$\alpha$ profile shape and LyC escape in the KLCS will test this formalism, which relies on a correlation between \fesc{} and Ly$\alpha$ peak separation that currently lacks direct observational support at high redshift. 

Both the \citet{chisholmFarUltravioletContinuumSlope2022} and \citet{mattheeReSolvingReionization2022} models highlight important existing relationships found between \fesc{} and galaxy property in our analyses, and present ionizing emissivities that both overcome recombination in the IGM at $z\sim8$ and avoid overproducing ionizing photons at $z<6$.
The trends between \fesc{} and galaxy properties presented in this work are vital for anchoring assumptions of \fesc{} during the epoch of reionization, where direct constraints on \fesc{} are impossible. Future reionization models can utilize these relationships to ensure consistency between reionization-era \fesc{} prescriptions and our empirical results, particularly in comparable galaxy populations that have similar luminosities and masses to those of our sample. We will extend our analysis of \fesc{} and galaxy properties to lower \luv{} in future work, which will elucidate most fundamental predictors of \fesc{} for a larger dynamic range of galaxy properties. 

\section{Summary} \label{sec:summary}

In this work, we examine the underlying processes behind the escape of ionizing radiation by exploring trends between \fesc{} and galaxy properties at \ziii{}. We accomplish this goal by leveraging multi-band photometry of galaxies observed spectroscopically as part of KLCS. We examined a subsample of 96 KLCS galaxies with photometry suitable for SED fitting, and determined galaxy population parameters of \mstar{}, SFR, sSFR, E(B-V), and age from these stellar-population synthesis fits. For each galaxy property, we sorted the 96 galaxies and divided them into three equal-sized bins, constructing a rest-UV composite spectrum for each bin. The main results regarding the estimated Lyman-continuum escape parameters of \fout{} and \fesc{} and their relationships with galaxy properties are as follows:

\begin{enumerate}
	\item We find significant correlations between \fesc{} and \wlya{} and anti-correlations between \fesc{} and \luv{} in the KLCS SED subsample, indicating that our sample is representative of the full KLCS and appropriate for constraining \fesc{} as a function of other galaxy properties \citep{pahlSearchingConnectionIonizingphoton2022}.
	\item We find significant anti-correlation between \fesc{} and E(B-V) across three bins of increasing E(B-V), although no correlation between \fout{} and E(B-V). The \fesc{} result indicates that dust modulates escaping ionizing radiation at \ziii{}. Such modulation naturally arises due to the spatial coincidence of neutral-phase gas and dust \citep{reddyConnectionReddeningGas2016,duRedshiftEvolutionRestUV2018,pahlRedshiftEvolutionRestUV2020}, and the fact that dust directly absorbs LyC photons \citep{reddySPECTROSCOPICMEASUREMENTSFARULTRAVIOLET2016}. These results are broadly consistent with anti-correlations found between \fesc{} and UV spectral slope at $z\sim3.5$ \citep{begleyVANDELSSurveyMeasurement2022} and in the local universe \citep{fluryLowredshiftLymanContinuum2022,fluryLowredshiftLymanContinuum2022a,chisholmFarUltravioletContinuumSlope2022}.
	\item Both \fout{} and \fesc{} are significantly correlated with \mstar{} and SFR. Trends between \fesc{} and \mstar{} have also been suggested in other LyC surveys at high and low redshift \citep{begleyVANDELSSurveyMeasurement2022,fluryLowredshiftLymanContinuum2022,fluryLowredshiftLymanContinuum2022a}. The sense of the relationships we observe is consistent with recovered anti-correlation between \fesc{} and \mstar{} for log($\mstar/\msun$)>8 galaxies in cosmological simulations \citep{maNoMissingPhotons2020,kostyukIonizingPhotonProduction2022}.
	\item Some cosmological zoom-in simulations of reionization-era galaxies connect stellar feedback and favorable ISM/CGM conditions for LyC escape \citep[e.g.,][]{maNoMissingPhotons2020}, which would plausibly manifest as elevated estimates of \fesc{} in galaxies with shorter inferred stellar ages and higher sSFR. However, we find no correlation between \fout{} or \fesc{} and stellar age or sSFR, providing no direct observational support for the synchronization of recent bursts of star formation and the escape of ionizing photons at these masses. These trends are consistent with the absent or weak correlation found between \wlya{} and stellar age in earlier work \citep{duRedshiftEvolutionRestUV2018,pahlRedshiftEvolutionRestUV2020,reddyEffectsStellarPopulation2022}.
\end{enumerate}

To date, these results represent the most comprehensive exploration of \fesc{} and SED-modeled properties at high redshift, grounding assumptions of \fesc{} for galaxies in the reionization era. Significant unknowns still remain for \fesc{} and its dependencies in galaxies less luminous than those in our sample, particularly at \ziii{}. In future work, we will extend our examination of \fesc{} and galaxy properties down to lower UV luminosities. Additional indirect diagnostics of \fesc{} that have proven promising in the local Universe can also be tested at high redshift with the KLCS. Ongoing follow-up of the Ly$\alpha$ line profiles of KLCS galaxies will elucidate potential trends between \fesc{} and Ly$\alpha$ peak separation, and Keck/MOSFIRE spectra in hand for a substantial subset of the KLCS will enable an examination of the relationships between nebular emission-line properties and \fesc{}. Using the results summarized in this section in tandem with future analyses of the KLCS, we will attempt to offer a unified picture of escaping ionizing radiation at \ziii{}. This picture is vital for understanding the contribution of star-forming galaxies to reionization at earlier times.
\linebreak
\linebreak
We acknowledge support from NSF AAG grants 0606912, 0908805, 1313472, 2009313, 2009085, and 2009278.
Support for program HST-GO-15287.001 was provided by NASA through a grant from the Space Telescope Science Institute, which is operated by the Associations of Universities for Research in Astronomy, Incorporated, under NASA contract NAS5-26555. 
CS was supported in part by the Caltech/JPL President's and Director's program.
Based in part on observations obtained with MegaPrime/MegaCam, a joint project of CFHT and CEA/IRFU,
at the Canada-France-Hawaii Telescope (CFHT) which is operated by the National Research Council (NRC) of Canada,
the Institut National des Science de l'Univers of the Centre National de la Recherche Scientifique (CNRS) of France,
and the University of Hawaii. This work is based in part on data products produced at Terapix available at the
Canadian Astronomy Data Centre as part of the Canada-France-Hawaii Telescope Legacy Survey, a collaborative project of NRC and CNRS. 
This paper includes data gathered with the 6.5m Magellan Telescopes located at Las Campanas Observatory, Chile. We thank D. Kelson for the use of his FourCLift FourStar Reduction code and for his assistance with it.
We wish to extend special thanks to those of Hawaiian ancestry on
whose sacred mountain we are privileged to be guests. Without their generous hospitality, most
of the observations presented herein would not have been possible.

\section*{Data Availability Statement}
The {\it HST} data referenced in this article are publicly available
from the Mikulski Archive for Space Telescopes. The ground-based
data presented here will be shared on reasonable request to the
corresponding author.


\end{document}